\documentclass[12pt,preprint]{aastex}

\newcommand{\arcm}{\mbox{\ensuremath{^{\prime}}}}

\usepackage{natbib}
\usepackage{longtable,lscape}
\begin{document}

\title{The Massive Star Forming Region Cygnus~OB2. II. Integrated Stellar Properties and the Star Formation History}

\author{N.~J.~Wright, J.~J.~Drake,}
\affil{Harvard-Smithsonian Center for Astrophysics, 60 Garden Street, Cambridge, MA~02138, U.S.A.}
\email{nwright@head.cfa.harvard.edu}

\author{J.~E.~Drew}
\affil{Centre for Astronomy Research, Science and Technology Research Institute, University of Hertfordshire, Hatfield AL10~9AB, U.K.}

\and

\author{J.~S.~Vink}
\affil{Armagh Observatory, College Hill, Armagh BT61~9DG, Northern Ireland, U.K.}

\begin{abstract}

Cygnus~OB2 is the nearest example of a massive star forming region, containing over 50 O-type stars and hundreds of B-type stars. We have analysed the properties of young stars in two fields in Cyg~OB2 using the recently published deep catalogue of {\it Chandra} X-ray point sources with complementary optical and near-IR photometry. Our sample is complete to $\sim$1~M$_{\odot}$ (excluding A and B-type stars that do not emit X-rays), making this the deepest study of the stellar properties and star formation history in Cyg~OB2 to date. From \citet{sies00} isochrone fits to the near-IR color-magnitude diagram, we derive ages of $3.5^{+0.75}_{-1.0}$ and $5.25^{+1.5}_{-1.0}$~Myrs for sources in the two fields, both with considerable spreads around the pre-MS isochrones. The presence of a stellar population somewhat older than the present-day O-type stars, also fits in with the low fraction of sources with inner circumstellar disks (as traced by the $K$-band excess) that we find to be very low, but appropriate for a population of age $\sim$5~Myrs. We also find that the region lacks a population of highly embedded sources that is often observed in young star forming regions, suggesting star formation in the vicinity has declined. We measure the stellar mass functions in this limit and find a power-law slope of $\Gamma = -1.09 \pm 0.13$, in good agreement with the global mean value estimated by \citet{krou02}. A steepening of the slope at higher masses is observed and suggested as due to the presence of the previous generation of stars that have lost their most massive members. Finally, combining our mass function and an estimate of the radial density profile of the association suggests a total mass of Cyg~OB2 of $\sim 3 \times 10^4$~M$_{\odot}$, similar to that of many of our Galaxy's most massive star forming regions.

\end{abstract}


\section{Introduction}

Stars do not form individually but in groups, with sizes ranging from small star forming regions (SFRs) to the massive superstar clusters seen in merging galaxies \citep{lada03}. Our current understanding of star formation is mainly derived from observations of nearby small star forming regions such as Taurus or Ophiuchus. However the conditions of star formation vary considerably between these regions and the massive SFRs that contain hundreds to thousands of OB stars and millions of low-mass stars. High stellar densities, strong stellar winds from OB stars and a large UV flux are likely to influence the products of the star formation process such as the initial mass function (IMF), the binary fraction and the evolution of protostellar disks. Our understanding of the influence of environment on these products is currently poor, and needs to improve if we are to develop a global understanding of star formation that can relate the small-scale properties with the evolution of galaxies and the survival of planetary systems.

The majority of massive SFRs are found in external galaxies, or close to the center of our own Galaxy. Their rarity and great distances therefore make them difficult objects to study. The one exception to this is Cygnus~OB2, which, at a distance of only 1.45~kpc \citep{hans03} is the closest known massive SFR. Early surveys of the region revealed the presence of $\sim$300 OB stars \citep{redd67}, a number that has grown as deeper surveys have penetrated the high extinction in its direction \citep[e.g.][]{mass91,come02,hans03}. There are currently 65 known O-type stars in the region making it the largest known concentration of spectroscopically confirmed O-type stars.

From a near-IR photometric study \citet{knod00} estimated the region contained over 100 O-type stars, $\sim$2600 OB stars, and had a total mass of 4--10~$\times 10^4$~M$_{\odot}$. However, the complications of background subtraction in the Galactic plane, combined with strongly variable extinction in the vicinity of Cyg~OB2, give cause to doubt such an estimate. Furthermore the question of the extent of Cyg~OB2 is made more complex by the growing evidence for stars older than the 1-3~Myr age appropriate to the OB stars: \citet{drew08} found a concentration of 5-7~Myr old A stars in 1~sq.~deg just to the south of Cyg~OB2, while \citet{come08} identified an unclustered spread of evolved stars up to $\sim$10~Myrs old over a wider area, 2~degrees in radius. In this study we obtain further evidence that would favour the present-day OB stars as the products of the latest but most certainly not the only phase of star formation.

Recent deep photometric surveys of the Galactic plane in the optical and near-IR have penetrated much of the extinction toward the region and are advancing our understanding of this massive association \citep[e.g.][]{drew08,vink08}. The combination of data from these surveys with that at other wavelengths is vital for studying the star formation process, particularly in the dense and highly obscured regions of the Galactic plane, and one that this work makes particular use of.

One of the objectives of this work is to measure the IMF down to low masses. The IMF is one of the most important measurable quantities in star formation studies. Combined with the time-variable star formation rate, the IMF determines the evolution of stellar systems ranging from small SFRs to entire galaxies. Measurements of the IMF have shown it to be highly uniform at high and intermediate masses \citep{luhm00,mass98} with a power law slope above 1~M$_{\odot}$ with an index of $\Gamma = -1.35$ \citep{salp55} for intermediate mass stars, possibly steepening to $\Gamma = -1.7$ for the most massive stars \citep{scal86}. At lower masses the IMF is less well constrained, but appears to flatten below 0.5~M$_{\odot}$ and exhibits a `turnover' around 0.1~--~0.3~M$_{\odot}$, with less stars of the lowest masses \citep{krou01a}.

Deviations from this uniformity are becoming increasingly rare as deeper and more complete observations overturn previous measurements and the majority of remaining observations have been shown to be within the statistical uncertainties of the seemingly universal mean \citep[e.g.][]{krou02}. While some theoretical models predict a shallower IMF in regions of high density or pressure, only a small number of observations in the vicinity of the Galactic Center \citep[e.g.][]{stol06} or in starburst regions \citep[e.g.][]{smit01a} support this and more measurements of the IMF in massive SFRs are necessary to resolve this. The most recent measurement of the IMF in Cyg~OB2, due to \citet{kimi07}, suggests that at high masses it is in fact steeper than the canonical value (though they estimate their spectroscopic sample becomes incomplete at masses below $\sim$15~M$_{\odot}$). This further supports the need for a thorough and complete measurement of the IMF in Cyg~OB2 down to lower masses.

In this paper we present an analysis and discussion of the stellar properties of sources in Cygnus~OB2 taken from the X-ray derived catalogue presented by \citet[][hereafter Paper~1]{ wrig09a}. 
In Section~2 we discuss the source catalogue and the various cuts applied to it to remove foreground and background contaminants. 
In Section~3 we present the near-IR properties of these sources, in Section~4 we study their X-ray properties and in Section~5 we derive mass functions for the sources. 
Finally, in Section~6 we discuss the implications of our findings on our understanding of Cygnus~OB2 and compare it to other massive SFRs.

\section{Cyg~OB2 membership and foreground contamination}
\label{s-complete}

The location of Cygnus~OB2 in the Galactic Plane introduces the non-trivial issue of contamination from foreground and background sources. This problem is further complicated by the high and variable extinction toward and throughout the Cygnus region. There are many methods that have been used to select members of a star forming region while minimizing the contamination from the field population, but one of the most effective and efficient is to use X-ray emission as a tracer of youth, since young pre-MS stars are orders of magnitude more luminous in X-rays than their MS equivalents \citep{prei05}. In Paper~1 a catalogue of 1696 sources X-ray sources was presented, extracted from {\it Chandra} observations of two fields in the center of Cyg~OB2. The catalogue also included optical and near-IR associations taken from the IPHAS \citep[INT Photometric H$\alpha$ Survey,][]{drew05}, 2MASS \citep[Two Micron All Sky Survey,][]{skru06} and UKIDSS \citep[United Kingdom Infrared Deep Sky Survey,][]{luca08} photometric catalogues. This work is based on this catalogue and principally the 1501 sources with either optical or near-IR counterparts that represents the main stellar sample.

The completeness of this catalogue is dependent on both the X-ray detections and the associations at other wavelengths. In Section~\ref{s-xlf} we show, by comparison with deeper X-ray stellar catalogues, that the completeness limit for this catalogue is $\sim$1~M$_{\odot}$. In Paper~1 it was shown that the near-IR photometry combined from 2MASS and UKIDSS observations will allow pre-MS stars down to 0.2~M$_{\odot}$ to be detected at an extinction of $A_V \sim 7$. In the north-western field, where some UKIDSS data is lacking, it was estimated in Paper~1 that the 2MASS observations reach a pre-MS depth of 1~M$_{\odot}$ at $A_V \sim 7$, though the lower extinction in this field means the observations will be slightly deeper than this. 

\begin{figure*}
\begin{center}
\includegraphics[width=320pt, angle=270]{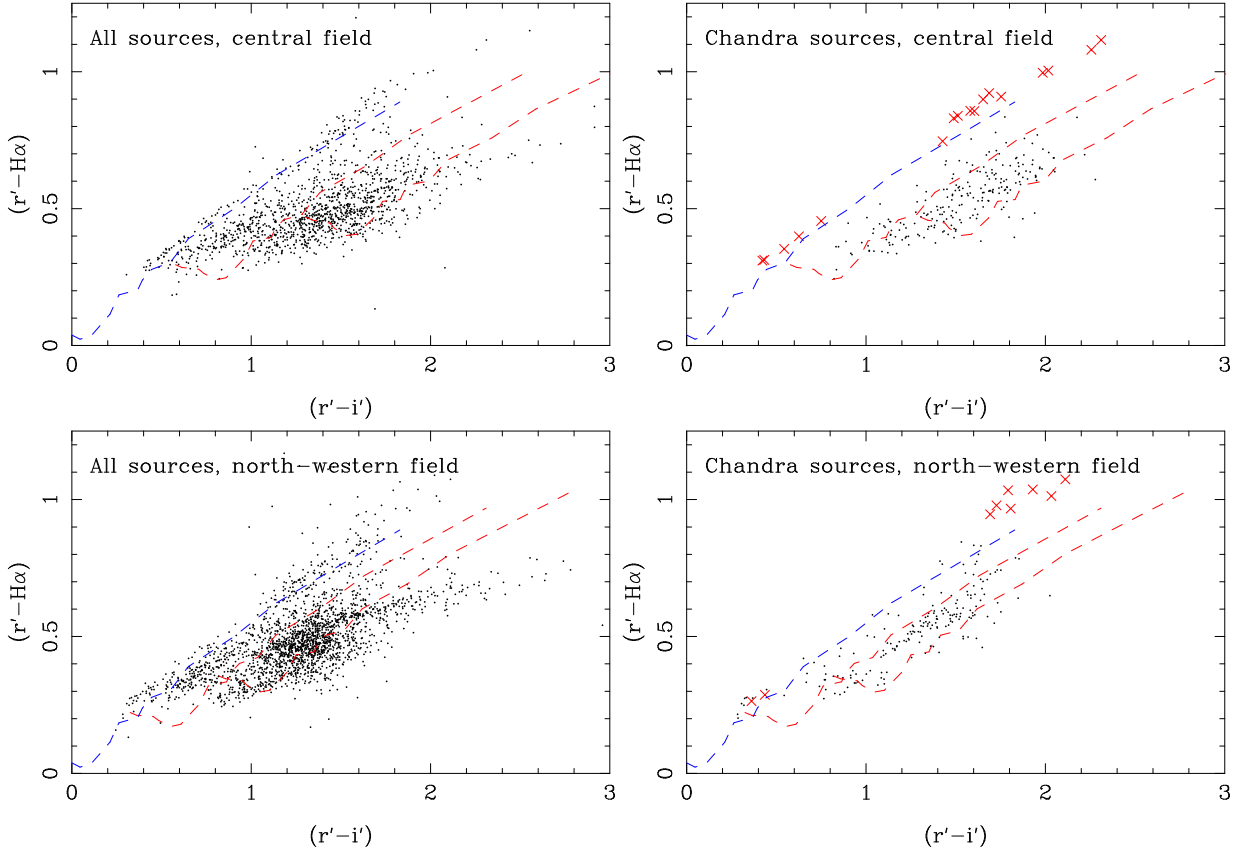}
\caption{IPHAS ($r' - $H$\alpha$, $r' - i'$) color-color diagrams for all IPHAS sources in the {\it Chandra} observational footprints (left-hand panels) and only for sources associated with {\it Chandra} sources (right-hand panels) for the central (upper panels) and north-western fields (lower panels). 
Only sources with $r' \leq 20$ and magnitude errors $\leq 0.1$ are shown. 
Blue dashed lines show unreddened main-sequences \citep{drew05}, while red dashed lines are the main-sequence tracks for reddenings of $A_V = 4.0$ and 7.0 (central field) and $A_V = 2.5$ and 5.0 (north-western field), illustrating the region approximately encompassing the Cygnus~OB2 population. Sources removed from the catalogue as foreground sources are shown as red crosses.}
\label{foreground_iphas_ccd}
\end{center}
\end{figure*}

Since not all young stars emit X-rays we must also be aware of the limits of X-ray emission mechanisms. O and early B-type stars (earlier than type B2, $>$10~M$_{\odot}$ are expected to generate X-rays in their radiatively-driven stellar winds \citep[e.g.][]{berg97}, while, at ages of $\sim$5~Myr, lower mass stars ($<$3.3~M$_{\odot}$) have substantial convective envelopes that produce X-ray emission through magnetic dynamo activity \citep[e.g.][]{pall81}. Intermediate mass stars do not have deep convective layers or strong stellar winds and are therefore not expected to emit X-rays. Studies that have observed X-ray emission from late B and A-type stars have attributed this to coronal emission from unresolved late-type companions \citep[e.g.][]{stel06}. 
The intermediate-mass completeness regime can be determined by counting the number of spectroscopically known OB stars in our field of view with X-ray detections. We find that the fraction of OB stars detected in X-rays decreases from 100\% for all O-type stars to 64\% at B0, 36\% at B1 and $\lesssim 20$\% for all later intermediate-mass stars. This suggests we are incomplete for the majority of B-type stars, and also most early A-type stars.

Finally, crowding is likely to have only a minor effect on completeness, because only the very central regions of the association are particularly crowded and these were observed close to on-axis with {\it Chandra} and have complete UKIDSS coverage, therefore allowing them the maximum possible spatial resolution in both X-rays and in the near-IR. 

\subsection{Removal of foreground sources}

While X-ray observations are excellent at separating young sources from older field stars, this method is not infallible since late-type main-sequence (MS) stars of all ages have been observed to have X-ray emission \citep[e.g.][]{vaia81}, albeit at generally lower levels. Nearby MS stars, although intrinsically fainter, could therefore have similar X-ray brightnesses as distant pre-MS stars, and thus be a potential contaminant in our catalogue. To remove any foreground sources, we employ one of the interesting features of the IPHAS ($r' - $H$\alpha$, $r' - i'$) color-color diagram (CCD) where the main sequence does not redden onto itself like it does in the near-IR CCD, but sweeps out an area in the color-color plane \citep{drew05}. This allows differently reddened stellar populations along the line of sight to be separated \citep[e.g.][]{drew08,sale09}.

In sightlines towards the Cygnus region, \citet{mccu71} found a high concentration of stellar sources between 200 and 300~pc and a strong decline in stellar density between 500~pc and 1~kpc, with the stellar populations making up Cygnus~OB2 and Cygnus~X beyond that. Figure~\ref{foreground_iphas_ccd} shows IPHAS CCDs for all IPHAS sources and for just those with {\it Chandra} X-ray counterparts in the areas covered by the {\it Chandra} observational footprints. It is clear from the CCD, as noted by \citet{drew08}, that the majority of foreground stars in the direction of Cyg~OB2 are nearby and only mildly reddened. The main Cyg~OB2 population, as traced by {\it Chandra}, can clearly be seen at between $A_V =$~4--7 in the central field and $A_V =$~2.5--5 in the north-western field.

It is possible that pre-MS members of Cyg~OB2 with H$\alpha$ emission may appear above the reddened main-sequences and closer to the foreground population (H$\alpha$ emission, frequently observed in young stars, causes the $r' - $H$\alpha$ color to increase). To preclude the removal of genuine members of Cyg~OB2 we compare the positions of candidate foreground stars in the near-IR color-magnitude diagram (CMD) with a foreground MS. Those sources whose positions in both diagrams were appropriate for pre-MS sources at the distance of Cyg~OB2 and potentially experiencing H$\alpha$ emission were retained. It should be noted that the majority of removed sources fall on a thin strip in the IPHAS CCD that would exhibit a larger spread if their positions in the CCD were due to the effects of H$\alpha$ emission.

For sources with only near-IR associations this method was not available to us and we had to rely on the positions of sources in the near-IR CMD. Due to the smaller effects of extinction in the near-IR and the fact that near-IR isochrones redden onto themselves it is not simple to separate the foreground population in a similar manner. Instead we may only identify sources with very blue colors as likely foreground sources and we have removed a small number of outlying sources in this way. In total we removed 46 sources ($\sim$3\%) from our catalogue as likely X-ray emitting foreground sources, which are listed in Table~\ref{t-masses}. These were predominantly removed from the 757 sources ($\sim$50\% of the original catalogue) with IPHAS photometry. Based on the fraction of IPHAS sources removed and the number of sources without IPHAS photometry we estimate that a small fraction $<$1\% of foreground contaminants may remain in the catalogue, but will be impossible to identify without spectra or further photometry. Finally we note that the spatial distribution of sources removed is uniform and does not exhibit a clustered distribution as the overall X-ray source distribution does (see Paper~1). We therefore confidently remove these sources from further consideration.

\section{Near-IR stellar properties}

We now consider the near-IR properties of the remaining 1497 {\it Chandra} sources with 2MASS or UKIDSS associations that we believe to be members of Cygnus~OB2. We will consider only those 2MASS sources with valid quality detections (photometric quality flags ``A" to ``D") and both 2MASS and UKIDSS sources with magnitude errors $<$1\%, in the relevant bands. Applying these restrictions to all three bands we are left with 1067 and 290 sources in the two fields, losing 50 (4\%) and 90 (24\%) sources in the central and north-western fields respectively. We lose a larger fraction of sources in the north-western field because the UKIDSS survey has yet to take data in this region and the fraction of 2MASS sources in this area with bad quality $K_s$-band photometric flags is high. However this should not affect the completeness limit of 1~M$_{\odot}$ since the available 2MASS photometry is deep enough to detect a one solar mass star at the distance and extinction of Cyg~OB2 (c.f. Paper~1). For ease of comparison between the two near-IR surveys and previous work in the literature all UKIDSS photometry was converted onto the 2MASS photometric system in Paper~1 using the conversions given by \citet{hewe06}. These conversions were used instead of the more accurate extinction-dependent transformations presented by \citet{luca08} because of the inhomogeneous extinction across the region (the expected systematic difference in the use of these transformations will be $\sim$0.03 mag at $J$, but only $\sim$0.01 mag at $H$ and $K$).

\subsection{Near-IR color-magnitude diagram}
\label{s-cmd}

\begin{figure*}
\begin{center}
\includegraphics[width=260pt, angle=270]{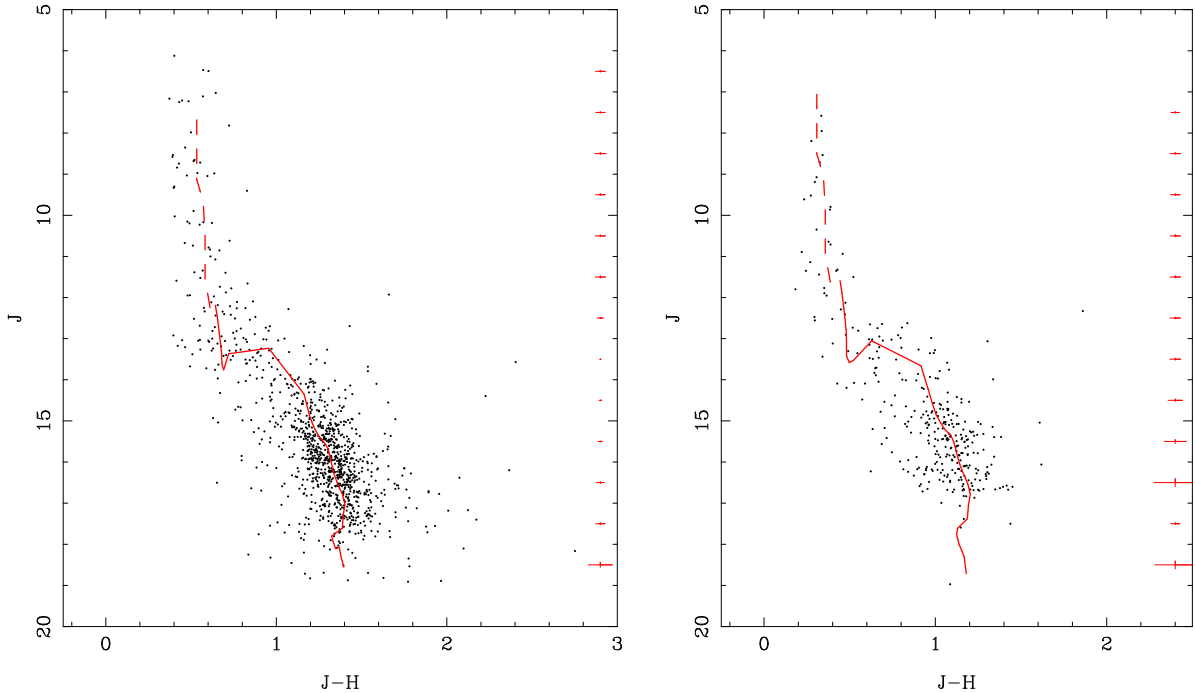}
\caption{Near-IR CMDs for Cygnus~OB2 sources in the central (left) and north-western (right) fields for all sources that meet the photometric requirements described above. 
Pre-MS isochrones \citep[from][]{sies00} are shown using the best fit parameters described in the text (3.5~Myrs, $A_V = 7.5$ for the central field and 5.25~Myrs, $A_V = 5.5$ for the north-western field), and at a distance of 1.45~kpc \citep[DM = 10.8,][]{hans03}. 
For masses above 8~M$_{\odot}$ we use absolute magnitudes from \citet{mart05} and \citet{cram97} and near-IR colors from \citet{koor83} and \citet{keny95}, for the O- and B-type stars respectively.
The MS and pre-MS were converted onto the 2MASS system using the data and relations given by \citet{keny95} and \citet{bess05} and then reddened using a standard Galactic extinction law \citep{howa83} and relative extinction relations given by \citet{schl98} and \citet{hans03}. 
Mean photometric errors for each magnitude interval are shown as red crossed along the right-hand side of each figure.}
\label{nearir_cmd}
\end{center}
\end{figure*}

Figure~\ref{nearir_cmd} shows ($J$, $J-H$) CMDs for sources in the two fields. We use the $J-H$ color instead of the $H-K$ color because it is less affected by circumstellar dust and has a smaller degeneracy between color and magnitude for pre-MS isochrones. Applying the photometric criteria discussed above to the $J$- and $H$-band photometry, there are 1076 and 325 valid sources, 96\% and 86\% of sources in each field. Both CMDs show a clear sequence with large numbers of sources down to $J \sim 19$. The greater depth in the central field is due primarily to the longer exposure of the {\it Chandra} observation, although the completeness of the UKIDSS survey in this field is also a benefit. The sources in the central field are also redder, as expected considering the greater extinction in this direction.

To determine stellar properties of the population of X-ray sources we have attempted to fit a number of pre-MS isochrones to our data, as shown in Figure~\ref{nearir_cmd}. We used \citet{sies00} isochrones for late spectral types (M~$< 8$~M$_{\odot}$) converted onto the 2MASS system using the data given by \citet{keny95} and \citet{bess05}. For earlier types we used the MS absolute magnitudes in \citet{cram97} and \citet{mart05} and colors from \citet{koor83} and \citet{keny95}. Goodness-of-fit tests, weighting each source using its color and magnitude errors, were performed to estimate the best-fitting isochrone and mean extinction of the sources in each field. Fits were attempted at a range of ages (at intervals of 0.25~Myrs) and reddenings (at intervals in $A_V$ of 0.25). To assess the validity of these results, 95\% confidence intervals were calculated for all fitted ages and extinctions using a bootstrapping technique in which our dataset was resampled 1000 times and refit. The range of fitted values provides a reliable estimate of the confidence intervals on each fit. Both the age and extinction were left as free parameters during the refitting.

We note first that, regardless of age, the lower mass stars are better fit by a slightly redder isochrone than the higher mass stars. This effect is most prominent for stars $< 2$~M$_{\odot}$, and is apparent in both fields, although it is stronger in the north-western field. 
This could be attributed to inaccuracies in the isochrones or near-IR colors for massive stars (particularly since a set of empirical 2MASS colours for massive stars does not exist in the literature), but we note that such a discrepancy has not been noted for other star forming regions. 
The use of two different sources for the near-IR photometry raises the question of whether the different photometric systems may have caused this discrepancy, but since the discrepancy is apparent in both fields and the north-western field uses primarily 2MASS photometry, the different photometric catalogues are unlikely to be responsible. The effect is also visible in the CMD presented by \citet[][their Figure~4]{alba07}, who used only 2MASS photometry in their analysis of the central {\it Chandra} field. 
The origin of this discrepancy is unclear. The high and variable extinction toward Cyg~OB2 and complex star formation history (see discussion in Section~\ref{discuss-age}) makes this difficult to resolve. Despite the apparent difference in reddening however, we show below that the difference between the fitted extinctions of the high- and low-mass populations are not significant.

Because of this we fit the high- (8-20~M$_{\odot}$) and low-mass ($<$8~M$_{\odot}$) populations separately. The extinction was allowed to vary for both fits, while the age of the region was only varied for the low-mass population because the high-mass population should already be on the MS and therefore, excluding the highest mass stars which may have evolved off the MS and which we do not include in our fit, are insensitive to age variations.
In the central field the high-mass stars are best fit at a reddening of $A_V \sim 7.25^{+0.5}_{-0.5}$, with the lower-mass stars redder with $A_V \sim 7.5^{+0.5}_{-0.75}$. In the north-western field the high-mass stars are fit with $A_V \sim 4.75^{+0.5}_{-0.75}$ with the low-mass stars having $A_V \sim 5.5^{+1.5}_{-1.0}$. Since there is no degeneracy between extinction and assumed distance, these values are independent of our choice of distance modulus. Compared to the mean extinctions calculated in Paper~1 from X-ray spectral fits these values are slightly larger, though within the uncertainties calculated here.

When fitting the low-mass component with pre-MS isochrones we attempted fits at both the older distance of 1.8~kpc \citep[DM=11.2,][]{mass91} and the revised distance of 1.45~kpc (DM=10.8) estimated by \citet{hans03}. Using DM=11.2 the low-mass stars are best fit with a $2.5^{+1.0}_{-1.25}$~Myr pre-MS isochrone, in agreement with previous studies based on the evolutionary state of the most massive stars in Cyg~OB2 \citep[e.g.][]{hans03}. At DM=10.8 the population is better fit with a slightly older 3.5$^{+0.75}_{-1.0}$~Myr isochrone (as shown in Figure~\ref{nearir_cmd}). This is older than the upper limit determined by \citet{hans03} based on main-sequence O6 and O7 dwarfs, but within the 95\% confidence interval.

In the north-western field we find consistently older ages for the stellar populations. At DM=11.2 the low-mass population is best fit at an age of $4.0^{+1.75}_{-1.25}$~Myrs and at DM=10.8 at an age of $5.25^{+1.5}_{-1.0}$~Myrs. For both fields the use of the closer distance proposed by \citet{hans03} requires an older pre-MS isochrone to fit the data. These ages will be discussed in more detail in Section~\ref{discuss-age}. 

\subsection{Near-IR color-color diagram}
\label{s-ccd}

\begin{figure*}
\begin{center}
\includegraphics[width=275pt, angle=270]{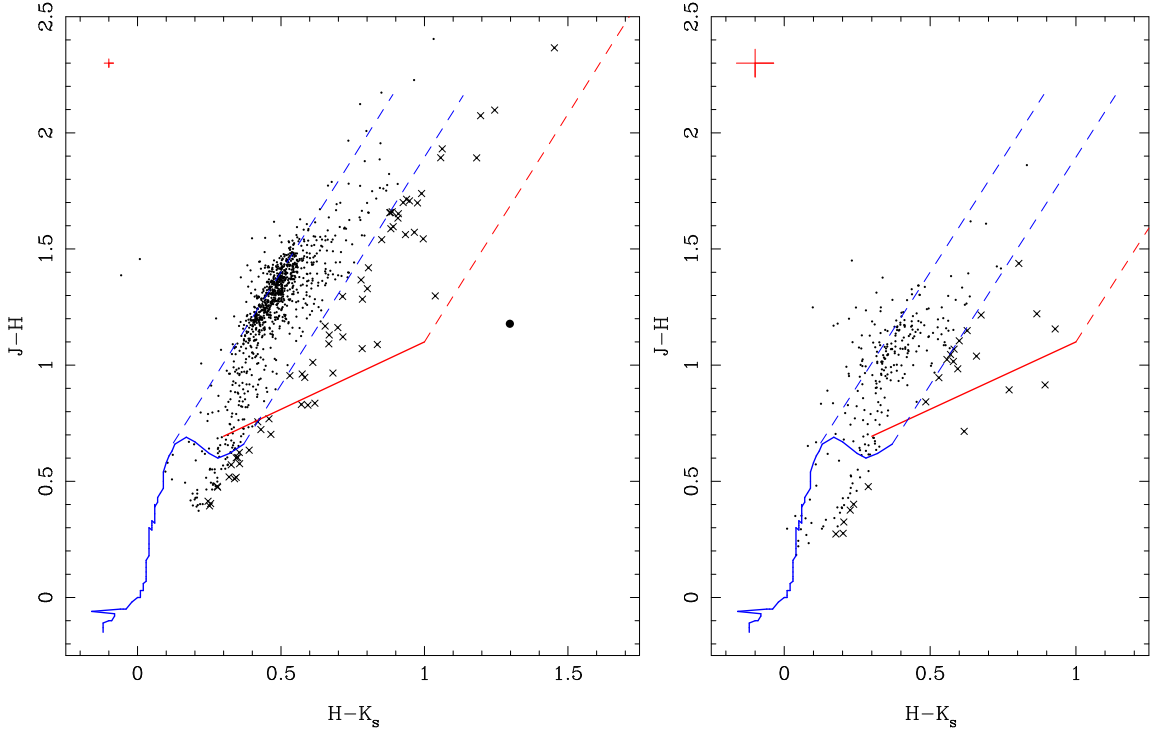}
\caption{Near-IR CCDs for Cygnus~OB2 sources in the central (left) and north-western (right) fields for all sources that meet the photometric requirements described above. 
The \citet{keny95} main-sequence is shown as a blue line, transformed onto the 2MASS photometric system using the relations given by \citet{bess05}. 
Reddening vectors with length $A_V = 15$~mag are shown extending from the MS at types K7V (M~$\sim$~0.6~M$_{\odot}$) and M6V (M~$\sim$~0.2~M$_{\odot}$) using the reddening relations described in Figure~\ref{nearir_cmd}. 
The classical T-Tauri star locus of \citet{meye97} is shown as a straight red line, also with a reddening vector extending from it, and sources falling within this region are marked as crosses. The highly reddened object in the central field is shown as a large dot. 
The mean photometric errors on the two colors is shown in the top left corner of each figure in red.}
\label{nearir_ccd}
\end{center}
\end{figure*}

Figure~\ref{nearir_ccd} shows the ($J-H$, $H-K_s$) CCD for the 1357 sources that meet our photometric requirements (fewer stars than present in the CMD in Figure~\ref{nearir_cmd}, on account of requiring satisfactory $K_s$ magnitudes also). For reference we also show the MS track from \citet{keny95} and the classical T-Tauri star (CTTS) locus from \citet{meye97}, both with reddening vectors extending from them. As noted above the majority of sources trace a reddened main-sequence, with a number of sources significantly more reddened still. The higher photometric errors of the 2MASS-dominated north-western field photometry is also evident when comparing the two fields and from the mean color error shown in each figure. In the central field there are two sources with large $J-H$ colors and low $H-K_s$ colors that do not correspond to plausible stellar colors and as such may have spurious photometry.

The near-IR CCD is often useful for estimating the evolutionary class of young stellar objects because sources with circumstellar material will be shifted to higher $H-K_s$ colors due to near-IR excesses in the $K_s$ band. These are Class~{\sc ii} objects \citep{lada87}, such as T-Tauri or Herbig Ae/Be stars, which are pre-MS stars with circumstellar accretion disks that are less evolved than the Class~{\sc iii} objects that have lost their inner circumstellar disks and clearly dominate our sample. We find 63 and 23 sources that fall within this region of the color-color plane, 5.9\% and 7.9\% of our catalogue for the central and north-western fields respectively. The fraction for the central field is slightly larger than the 4.4\% determined by \citet{alba07}, which can likely be attributed to our use of UKIDSS data to probe lower-mass stars that retain circumstellar disks for longer periods of time \citep[e.g.][]{hill98,hais01a}. To the right of the reddened CTTS is a region typically occupied by less evolved Class~{\sc i} protostars, or evolved stars, with significant amounts of circumstellar material. We find only one object in this region, and from its isolation and lack of IPHAS photometry we are unable to determine whether it is a true member of Cyg~OB2 or a background source.

The fractions of Class~{\sc i} and {\sc ii} sources are significantly lower than that found from studies of other similar age star forming regions in the near-IR such as M17 \citep[37\%,][]{hoff08} and the Orion Nebula Cluster \citep[$>$55\%,][]{hill98}\footnote{We note that while different studies often use different criteria for defining near-IR excesses and quote excess fractions over different stellar mass ranges, the clearly higher fraction of excess sources in these regions compared to Cyg~OB2 is worthy of attention and explanation}. \citet{alba07} suggested that the low fraction of circumstellar disks in Cyg~OB2 was due to the strong UV radiation field from the large number of OB stars in the vicinity that causes photoevaporation of the disk material. Although predicted by theory \citep[e.g.][]{john98}, an inverse relationship between disk fractions and UV radiation field has only been mildly observed in star forming regions to date \citep[e.g.][]{balo07,guar07}. To a certain extent our results support such a theory, with the fraction of circumstellar disks lower in the center of the star forming region where the UV radiation field from OB stars is stronger. However the difference is both statistically small and this interpretation over-simplifies the dynamical and 3-dimensional structure of the region \citep[e.g.][]{balo07} and therefore does not allow an accurate comparison between disk fractions and the UV radiation field. However, If Cyg~OB2 contains a significant fraction of sources with ages $\geq 5$~Myr, then the disk fractions of 5~--~10\% found here are perfectly reasonable \citep[e.g.][]{hill05}. Given the growing evidence for an older generation of star formation in the region (see discussion in Section~\ref{discuss-age}) we suggest that this is a more likely explanation for the low fraction of circumstellar disks than invoking UV photoevaporation from OB stars.

The use of near-IR colors is however an imprecise method of tracing the population of disk bearing stars since a $K_s$-band excess will only reveal hot dust from the inner parts of a circumstellar disk. The combination of near- and mid-IR photometry is necessary to produce a more complete picture of the evolution of circumstellar disks in a star forming region. Forthcoming observations of the Cygnus region from the {\it Spitzer Space Telescope} will allow this matter to be probed in more detail.

\section{X-ray stellar properties}

In this section we discuss how the integrated X-ray properties of sources in Cyg~OB2 may be used to study the star forming region as a whole, and make comparisons with the properties of X-ray sources in other star forming regions.

\subsection{Median X-ray photon energy}
\label{s-median}

\begin{figure}
\begin{center}
\includegraphics[width=130pt, angle=270]{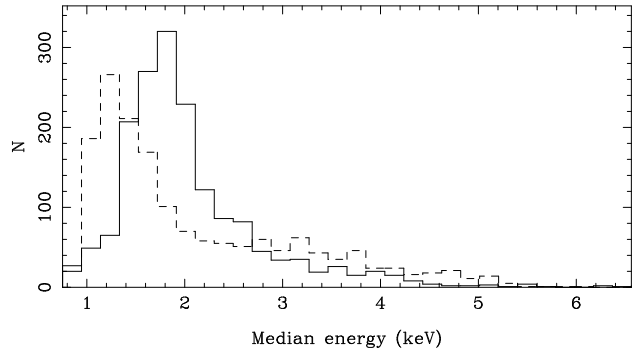}
\caption{The distribution of median energies of the Cygnus~OB2 {\it Chandra} sources (full line) compared to the Orion Nebula Cluster \citep[][dashed line]{getm05a}.}
\label{medianenergy}
\end{center}
\end{figure}

\citet{feig05} showed that the median energy of X-ray sources is a reliable indicator for the absorbing column density, log~$N_H$, obtained from spectral fitting procedures. In Figure~\ref{medianenergy} we show the distribution of median energies of {\it Chandra} sources in Cygnus~OB2 compared to that of the well-studied {\it Chandra} Ultra-deep Orion Project \citep[COUP,][]{getm05a} of the Orion Nebula Cluster (ONC). The peak median energy of Cygnus~OB2 sources is higher than that of Orion because of the higher extinction toward Cyg~OB2 compared to Orion.

The COUP sample also shows a smaller population with a higher median energy around 3~keV that is indicative of a population of embedded sources with higher extinction. However, such a second population is not clearly visible in the Cyg~OB2 distribution with the high median energy sources appearing to constitute the tail-end of the standard distribution. This can be quantified as the fraction of sources in the two catalogues that would be classed as ``heavily obscured'' by some authors, i.e. with $\langle E \rangle \geq 3.0$~keV. In Cyg~OB2 only 10\% of the 1696 sources fall in this class, compared to 21\% of Orion sources. This relative lack of embedded sources is observed in both {\it Chandra} fields. It is also not a property of the typically higher mass of our Cyg~OB2 sources compared to those in the deeper low-mass COUP sample: filtering both samples on X-ray luminosity (which correlates well with stellar mass) and comparing only those sources with $L_X > 29.5$~erg~s$^{-1}$ (our approximate X-ray completeness limit, see Section~\ref{s-xlf}) also shows a deficit of embedded Cyg~OB2 sources. 

While it appears there is a moderate evolution in stellar X-ray luminosity with age \citep[$L_X$ decreasing with age, but $L_X / L_{bol}$ increasing,][]{prei05}, and there might also be an associated change in spectral hardness as stars age, we expect such effects to be very small compared to the gross changes observed in median photon energy as a result of the different absorption characteristics of the ONC and Cygnus~OB2 regions. We conclude that the relative lack of sources with high median X-ray energies indicates that, unlike the ONC, the central region of Cyg~OB2 does not contain an embedded population of recently formed stars and that star formation in the vicinity of Cyg~OB2 has declined significantly. 

\subsection{The X-ray luminosity function}
\label{s-xlf}

Different young star forming regions, such as Cepheus~OB3 \citep{getm06}, M17 \citep{broo07}, and NGC~6357 \citep{wang07}, have all been observed to have X-ray luminosity functions (XLFs) that exhibit similar lognormal shapes. Since the XLF is effectively a convolution of the IMF and the X-ray luminosity -- mass relation for pre-MS stars, it may be used to study the properties of a stellar population without making any assumptions about the region or the objects being studied (e.g. by using pre-MS evolution models). Furthermore, if we assume that the IMF is invariant in young clusters, at least as far as can be measured, the completeness of our sample may be estimated by comparing the XLF with that determined from deeper observations of other clusters, such as the COUP study which is complete to $\sim$0.1~M$_{\odot}$.

\begin{figure}
\includegraphics[width=185pt, angle=270]{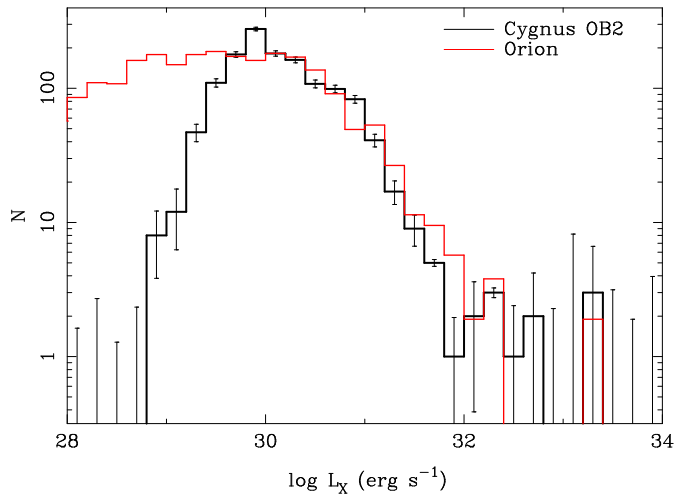}
\caption{The extinction-corrected full-band (0.5~--~8.0~keV) X-ray luminosity function for all ``lightly obscured'' ($\langle E \rangle \leq 3.0$~keV) {\it Chandra} sources in Cygnus~OB2 (black line) compared to the equivalent XLF from the {\it Chandra} Orion Ultradeep Project \citep[COUP, red line,][]{getm05}, which has been adjusted to the revised distance of 400~pc for the ONC \citep{muen08}. The COUP XLF has been renormalized to fit the high-mass Cyg~OB2 XLF. All Cyg~OB2 X-ray luminosities are scaled to a distance of 1.45~kpc. XLF 1$\sigma$ error bars for the Cyg~OB2 sample are also shown.}
\label{xlf}
\end{figure}

Figure~\ref{xlf} shows a comparison between the XLF of lightly obscured ($\langle E \rangle \leq 3.0$~keV) Cygnus~OB2 sources and those from the well-studied Orion Nebula (we show only the lightly obscured populations since Cyg~OB2 has a smaller heavily obscured population compared to the ONC, and this will influence the XLF). Error bars for the Cyg~OB2 XLF showing the 1$\sigma$ confidence level were estimated from Monte Carlo simulations of the XLF, where individual X-ray luminosities were drawn from a Gaussian distribution with mean equal to the measured X-ray luminosity and variances based on the X-ray luminosity error. The COUP XLF has been renormalized to fit the Cyg~OB2 XLF at the intermediate- to high-mass end.

Between log~$L_X \sim 29.5$~--~31.5~erg~s$^{-1}$ the ONC and Cyg~OB2 XLFs are in good agreement. At lower luminosities the two XLFs differ, with fewer Cyg~OB2 X-ray sources compared to the ONC. This represents our completeness limit, from which we determine a mass limit of 1~M$_{\odot}$ using the X-ray luminosity to mass correlation in \citet{prei05a}.\footnote{Note that an X-ray luminosity of log~$L_X \sim 29.5$ corresponds to a mass of $\sim$0.25~M$_{\odot}$ using the conversion for \citet{sies00} models presented by \citet{prei05a}, but since the X-ray luminosity -- mass relation has a spread of at least 1~dex, we estimate the actual completeness using the range in luminosity in Figure~3 of \citet{prei05a}.}. Beyond this our completeness drops off rapidly towards lower masses with our sample only 50\% complete at $0.75 M_{\odot}$ and $< 20$\% complete at $0.5 M_{\odot}$. Considered separately the two fields show similar XLFs down a similar completeness limit, despite an exposure difference of a factor of two. This is a product of the $L_X$~--~mass relation which is level between 1~--~10~M$_{\odot}$ and then drops rapidly at lower masses with a high dispersion \citep{prei05a}. Beyond the completeness limit, the XLF falls off quicker in the north-western field than it does in the central field due to the different in X-ray exposure and the lack of 

\section{Stellar masses and mass functions}
\label{masses}

The compilation of a large sample of cluster members with near-IR photometry and fitted pre-MS isochrones allows stellar masses to be estimated for a large and relatively complete sample. This then allows the mass function of sources in Cyg~OB2 to be determined.

Individual stellar masses were calculated based on the positions of the sources in the ($J$, $J-H$) CMD by tracing an extinction vector back to the theoretical MS or pre-MS isochrones described in Section~\ref{s-cmd}. For this purpose we assume a distance to Cyg~OB2 of 1.45~kpc \cite{hans03}. We will initially assume that all stars in each field were born coevally, ignoring any possible age spread, therefore allowing a direct relationship between the position of the source in the CMD and its mass and extinction. For this we will use the best-fitting ages determined in Section~\ref{s-cmd} for DM~=~10.8. Because of the form of the pre-MS isochrone in the near-IR CMD a number of sources have degenerate solutions for their masses. This degeneracy was removed by dereddening these sources using their X-ray spectral-fitted hydrogen column densities and the relation $N_H = 2.2 \times 10^{21} A_V$~cm$^{-2}$ \citep[e.g.][]{gore75,ryte96} and then finding the closest solution on their reddening locus. Table~\ref{t-masses} lists the masses and extinctions for all sources in Cyg~OB2 with near-IR photometry determined using this method.

Masses derived using this method will be dependent on a number of factors. The $J$-band magnitude errors, as shown in Figure~\ref{nearir_cmd}, are small, and will therefore produce only a small error in the derived mass (at $J \sim 17$, where $M = 0.65 M_{\odot}$, the typical $J$-band error of 0.024~mags corresponds to an error of only 0.02~M$_{\odot}$). Other photometric ``discrepancies'', such as treating a binary star as a single star or not considering the effects of variability will affect the derived stellar mass slightly more. The influence of circumstellar material on the stellar masses is reduced by our use of the $J-H$ color as opposed to the $H-K_s$ color in deriving masses from near-IR colors. The age of the region, estimated in Section~\ref{s-cmd}, will also affect the derived masses, and this is discussed later. Finally, the choice of pre-MS isochrones could potentially have a significant effect. Studies show that using different isochrones can produce mass differences of up to a factor two \citep{hill04}, though this effect is strongest at low masses.

\begin{table}
\caption{Derived stellar properties of sources in Cygnus OB2 with optical or near-IR associations. Only the first 10 rows are shown. The complete table is available in the electronic edition of the journal.}
\label{t-masses}
\begin{tabular}{rccc ccc}
\tableline 
No. & RA & Dec & Member & Mass & A$_V$ & Source \\ 
&&&& (M$_{\odot}$) \\
\tableline 
1 & 20:31:16.03 & 41:26:32.2 & Y & 1.9 & 10.0 & CMD \\ 
2 & 20:31:19.32 & 41:29:32.3 & Y & 5.7 & 8.9 & CMD \\ 
3 & 20:31:20.60 & 41:31:51.1 & Y & 5.2 & 7.7 & CMD \\ 
4 & 20:31:21.15 & 41:29:04.1 & Y & 0.66 & 6.7 & CMD \\ 
5 & 20:31:22.14 & 41:28:40.3 & Y & 1.8 & 2.2 & CMD \\ 
6 & 20:31:22.28 & 41:29:29.9 & Y & 1.6 & 9.0 & CMD \\ 
7 & 20:31:23.32 & 41:31:50.7 & Y & 1.4 & 3.1 & CMD \\ 
8 & 20:31:23.47 & 41:27:15.1 & Y & 1.7 & 8.1 & CMD \\ 
9 & 20:31:23.56 & 41:29:48.9 & Y & 1.9 & 2.8 & CMD \\ 
10 & 20:31:23.63 & 41:30:40.7 & Y & 1.3 & 5.4 & CMD \\ 
\tableline 
\end{tabular} 
\newline Notes. Col. (1): X-ray source number. Cols. (2)-(3): Position of optical or near-IR source that X-ray source was matched to. Col. (4): Association membership information (Y/N). Col. (5): Stellar mass. Col. (6): Extinction toward source. Col. (7): Method by which mass and extinction were determined. For sources with spectral information in the literature a reference is provided, while sources where masses were determined from the CMD are listed as ``CMD".
\end{table} 

For many of the most massive stars in Cyg~OB2 a considerable body of spectroscopic observations exists that may be used to derive spectral types and masses. In Paper~1, 54 sources with spectroscopic observations in the literature were associated with {\it Chandra} sources of which 51 have full spectral types. We determined zero-age masses for all these sources using the zero-age masses from the published parameters of \citet{mart05} for O-type stars and from the collated information presented by \citet{kimi07} for B-type stars (primarily from stellar evolutionary models).

\subsection{The mass function of Cygnus OB2}

The mass functions (MFs) of very young SFRs are often described as initial mass functions, although this is rarely true. The IMF may be disrupted within as little as $\sim$1-2~Myr due to the ejection of members through interactions and the demise of the highest mass members. Considering the ages estimated in Section~\ref{s-cmd} and evidence in the literature for ejected massive stars from Cyg~OB2 \citep[e.g.][]{come07}, it is important to make this distinction and understand its implications.

\begin{figure*}
\begin{center}
\includegraphics[width=180pt, angle=270]{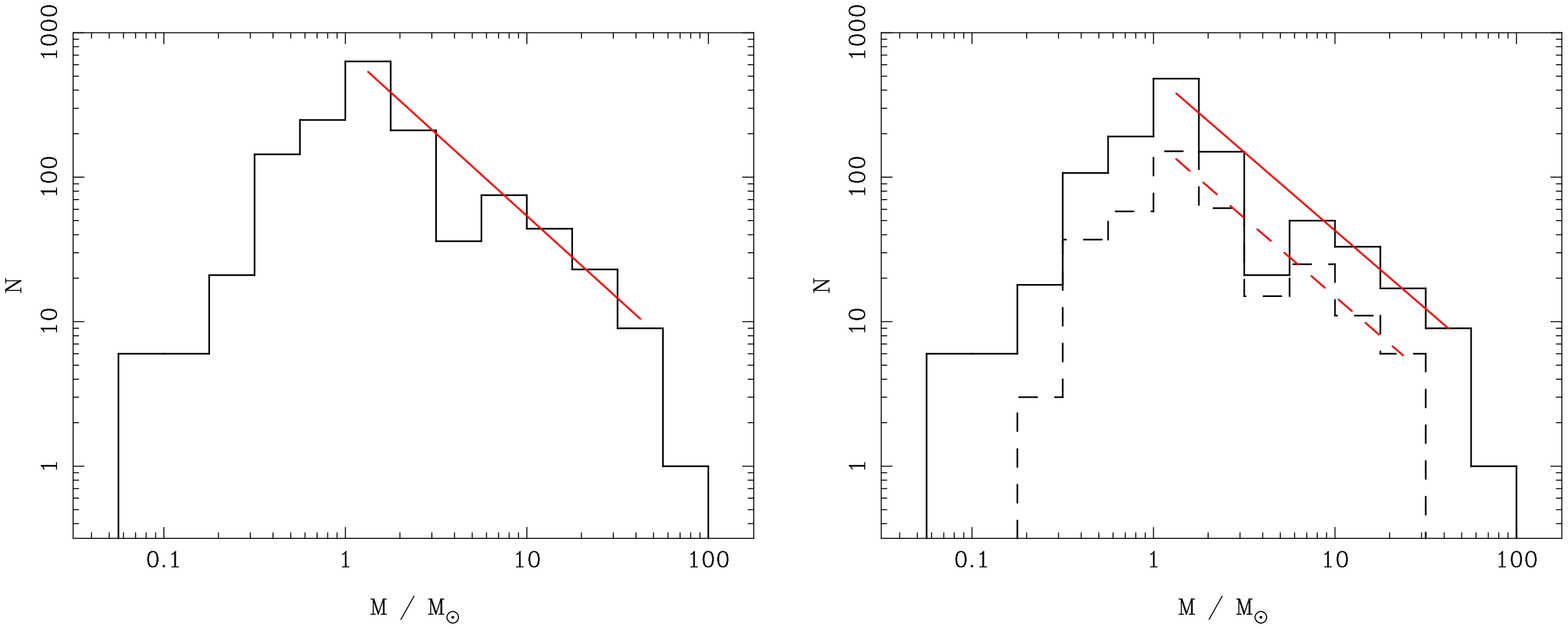}
\caption{Mass functions (MF) for all sources in Cygnus~OB2 with stellar masses derived from near-IR photometry and existing spectroscopic observations. {\it Left:} MF for both fields combined; {\it right:} MF for the central (full line) and north-western (dashed line) fields. $\chi^2$ fits to the MFs within the completeness limits ($M > 1 M_{\odot}$ and excluding early A and late B-type stars) are shown as red lines, with slopes of $\Gamma = -1.09 \pm 0.13$ (combined field), $\Gamma = -1.08 \pm 0.15$ (central field), and $\Gamma = -1.09 \pm 0.10$ (north-western field).}
\label{imf}
\end{center}
\end{figure*}

Figure~\ref{imf} shows mass functions for both fields in Cyg~OB2 and for the combined fields, all of which reach a maximum at $\sim$1~M$_{\odot}$, our estimated completeness limit (Section~\ref{s-xlf}). Excluding the bins spanning the range 3~--~10~M$_{\odot}$ that contain A and late B-type stars not expected to emit X-rays (see Section~\ref{s-complete} and our discussion of the completeness of our sample) the mass functions show a clear power law slope, steepening at higher masses. 
Excluding the most massive stars we find a power-law slope of $\Gamma = -1.09 \pm 0.13$ in the combined fields, in agreement with the value of $\Gamma = -1.3 \pm 0.3$ estimated by \citet{krou01a} from accumulated observations, and only slightly shallower than the canonical Salpeter value. 
When the two fields are considered separately we find negligibly different slopes of $\Gamma = -1.08 \pm 0.15$ in the central field and $\Gamma = -1.09 \pm 0.10$ in the north-western field. 
When considering only the O-type stars we find a much steeper slope with $\Gamma = -2.72 \pm 0.52$. 

Previous measurements of the MF\footnote{The majority of previous mass function measurements for Cyg~OB2 in the literature refer to the IMF, most likely because of the previously assumed age of $\sim$2~Myrs. A full understanding of the age and star formation history of the region is therefore clearly important for correctly interpreting the measurements of the mass function.} in Cyg~OB2 have mostly been derived from spectroscopy of the massive stars. \citet{mass91} measured $\Gamma = -1.0 \pm 0.3$ for all stars more massive than 9~M$_{\odot}$ and $\Gamma = -0.9 \pm 0.5$ using only the highest quality photometry and stars more massive than 15~M$_{\odot}$. Compared to our high-mass MF this is a much shallower slope, although the masses used were derived from older models for massive stars. 
More recently, \citet{kimi07} derived a slope of $\Gamma = -2.2 \pm 0.1$ for spectroscopically studied stars more massive than 15~M$_{\odot}$ across a wide area of Cyg~OB2, deriving masses from updated stellar evolutionary models and the revised O star stellar parameters presented by \citet{mart05}. Our high-mass MF is in good agreement with their value, mainly because of our use of similar models and stellar parameters. However, we have shown here that the MF at lower masses is not as steep and is closer to the canonical value. We suggest that the steep high-mass slope of the MF is not intrinsic but is a product of the older stellar populations in the region having lost their most massive members, thereby steepening the mass function.

The only previous measurement of the mass function at lower masses is due to \citet{knod00} who found $\Gamma = -1.6 \pm 0.1$ down to 2~M$_{\odot}$ from a photometric analysis of 2MASS observations in Cyg~OB2. However, his use of main-sequence magnitudes and colors for a pre-MS stellar population puts the derived masses into question. Our measurement is therefore the first deep measurement of the pre-MS mass function in Cygnus~OB2 and its agreement with the \citet{krou01a} value, supporting the view that the IMF is invariant.

\subsection{Effect of age uncertainties on mass function slope}

The uncertainty in our estimation of the age of Cyg~OB2 has repercussions for stellar masses derived from fitting pre-MS isochrones to the near-IR CMD. We also cannot assume that all the stars within a single field were formed coevally (see the discussion in Section~\ref{discuss-age}). To test the impact on the MF slope of this age uncertainty we altered the estimated age and regenerated the individual stellar masses and the MF. We find that when the estimated age is varied by $\pm1$~Myr the derived MF slope changes by less than 10\%, with the MF in the north-western field varying much less than in the central field. Even varying the age over a large range (by $\pm2$~Myr) only causes a MF slope variation of $\sim$20\% (for the central field) and $< 10$\% in the north-western field. Age variations larger than this cause much greater variations in the mass function slope, particularly in the central field, but are well outside the 95\% confidence intervals.

We also studied how the MF would vary if all the stars were not formed coevally but formed over a range of ages (an age spread that might explain the wide spread in the CMD, see discussion in Section~\ref{discuss-age}). Since the spread in the CMD could be caused by both a spread in extinction and a spread in ages we did this by selecting ages for each sources that would minimize the resulting spread in extinction. This is the opposite process to that used above where a single age was used, thereby minimizing the spread in ages (the true situation is likely to be somewhere between these two extremes). The resulting MF has a slope of $\Gamma = -1.03 \pm 0.11$ in the combined fields, $\Gamma = -0.98 \pm 0.13$ in the central field, and $\Gamma = -1.05 \pm 0.06$ in the north-western field. These small variations in the slope of the MF strengthen its validity despite the uncertainty over the age of the region.

\section{Discussion}

In this section we discuss the results obtained here in the context of our current understanding of Cygnus~OB2, its size, age, and star formation history. We also compare the derived properties of the region with those of the small but growing number of massive SFRs that have been studied.

\subsection{The age of Cygnus~OB2}
\label{discuss-age}

The main constraints we have on the age of Cyg~OB2 are derived from observations of its most massive stars. \citet{hans03} estimated an age of $2 \pm 1$~Myrs to accommodate the observed mix of main sequence and evolved OB stars: very young sources such as \#A37 from \citet{come02}, which \citet{hans03} classified as an O5V((f)), cannot be older than 1~Myr (based on evolutionary tracks); while supergiants such as Cyg~OB2 \#9, type O3If \citep{mass91}, are believed to be at least 3~Myrs old. The presence of a main sequence extending from type O6, which also cannot be older than 3~Mys, supports this age estimate.

There has been growing evidence for a large number of evolved supergiant stars in the region \citep{mass91,come02}. \citet{hans03} noted that as the search for OB stars in the region was extended to larger radii the fraction of evolved massive stars increased, leading her to suggest that our view of Cyg~OB2 is becoming contaminated by non-members. Many of the new evolved massive stars identified by \citet{come02} were found to the south of the original association in a region where \citet{drew08} have since uncovered a population of A-type dwarf stars with an age of 5~--~7~Myr. \citet{drew08} suggest that this region may represent a larger and older parent cluster.

In Section~\ref{s-cmd} we derived ages of $3.5^{+0.75}_{-1.0}$ and $5.25^{+1.5}_{-1.0}$~Myrs for the central and north-western fields, both with large spreads in the CMD that cannot solely be explained by photometric errors. Such wide spreads in CMDs have previously been interpreted as due to wide age spreads in association members \citep[e.g.][]{pall00}. However other authors have suggested that the observed CMD spreads are due to a mixture of photometric errors, variability in pre-MS stars \citep[up to several magnitudes for classical T-Tauri stars,][]{herb94}, binarity \citep{prei99}, episodic accretion phases \citep{bara09} or a spread in line-of-sight distances or extinctions. Many of these factors are not well understood, and in the case of Cyg~OB2 particularly the spread in extinction. \citet{sale09} measured the rise of extinction with distance towards Cyg~OB2 and found a substantial increase in extinction at $\sim$1.5~kpc, which they estimate could be confined to a depth of $\leq 100$~pc. If a significant spread of extinction did exist across Cyg~OB2, it could, combined with the effects of binarity and pre-MS variability, explain a large proportion of the spread in the CMD. Furthermore, given the uncertainties on the two ages derived in this work, the age difference between the two fields is not significant in itself and could fit with a picture of a single burst of star formation in the region.

However, combining the ages derived here with the existing evidence for very young ($\sim$1~Myr) and slightly older ($\sim$5~--~7~Myr) populations in Cyg~OB2 suggests a picture of a region with a large spread in ages. In such a picture the central field, for which we derived an age of $3.5^{+0.75}_{-1.0}$~Myr, merges with the young OB-star dominated association, while the north-western field is biased to a greater age. We cannot yet know to what extent there are two distinct generations of stars present, or whether there is a real age spread (with the youngest stars co-located with the OB stars).

\subsection{The size and mass of Cygnus OB2}

As shown in Section~\ref{s-complete}, the catalogue presented in Paper~1 is complete to 1~M$_{\odot}$ within the caveats of X-ray emitting stars. We can therefore combine the MF determined here and the stellar density within our completeness limits to estimate the total size and mass of the entire Cyg~OB2 association. For the purposes of this exercise we will consider the Cygnus~OB2 association to consist of all sources young enough to be detected in X-rays (i.e. $< 100$~Myrs). While this may include multiple generations of star formation that may have occurred on different spatial scales, until the different populations can be separated this is the only option available to us. Also, because our observations only sample two small fields in the center of the association, we must combine our results with previous estimates of the total spatial extent of the SFR available in the literature.

\citet{knod00} analysed the morphology of Cyg~OB2 by fitting a King profile \citep{king62} to 2MASS star count data, estimating it to be a spherically symmetric association of $\sim 2^{\circ}$ diameter. This analysis necessarily relied heavily on performing a background subtraction using the stellar content of a nearby field, despite the inherent difficulties involved in doing this along complicated sightlines in the Galactic plane. An improved morphological analysis may be made using the radial stellar density of {\it Chandra} sources in the direction of the north-western field where a radial length of 30\arcm\ is possible using the two fields (see Figure~1 in Paper~1). 

We find that the stellar density drops off much more quickly than estimated by \citet{knod00}, with approximately one third as many stars detected by us in the 1~--~2.5~M$_{\odot}$ mass range (where our data remain complete) at distances of $>$15\arcm, compared with the King profile determined by \citet{knod00}. Using our radial density profile and assuming that Cyg~OB2 is spherically symmetric we have estimated a new King profile for Cyg~OB2. Using our peak stellar density and the mass function slope estimated above, extrapolated down to 0.01~M$_{\odot}$ using the multi-stage power-law IMF from \citet{krou01a} we calculate a total stellar content of $5.7 \times 10^4$ stars and a total mass of $2.8 \times 10^4$~M$_{\odot}$. However it is clear from the spatial distribution of A-type dwarfs in Cyg~OB2 presented by \citet{drew08} that the structure of Cyg~OB2 is far from spherical. The authors note a distinct lack of A-type dwarfs to the east of Cyg~OB2 and a pronounced concentration towards the southern part of the cluster that may increase the size of the association. We therefore estimate a total stellar mass of $(3 \pm 1) \times 10^4$~M$_{\odot}$ for the entire association.

Using this estimated density profile and our derived MF we can estimate the expected number of stars of different masses in Cyg~OB2. A total mass of $3 \times 10^4$~M$_{\odot}$ implies a total of $\sim$1200 OB stars, approximately half that predicted by \citet{knod00}. Our current MF predicts a total number of O-type stars of $\sim$75, only slightly larger than the currently known number of 65 O-type stars in Cyg~OB2 \citep{mass91,hans03,kimi07,come08,negu08}. This may suggest that there are more O-type stars to be discovered in Cyg~OB2. If we assume that the high-mass MF slope has been influenced by the demise of the most massive stars in the region and that it was originally $\Gamma = -1.3$ \citep{mass98}, we estimate the number of stars that have evolved to their end states to be $\sim$12. There are no known supernova remnants in Cyg~OB2, although there is some evidence for stellar remnants: \citet{bedn03} and \citet{abdo09} identify two pulsars within Cyg~OB2, with ages $\sim 2 \times 10^4$~yr \citep{cami09}, that likely represent recently expired massive stars, while \citet{come07} identified an O4If runaway star that appears to have been ejected from Cyg~OB2 $\sim$1.6~Myr ago, possibly following the supernova explosion of a more massive binary companion. 

\subsection{Comparison with other massive star forming regions}

Table~\ref{msfrs} lists some of the properties of Cyg~OB2 estimated in this work along with those of a number of other massive star forming regions. With the exception of the ONC the majority of these regions have not been studied to as low a stellar mass, with properties determined in different ways by different authors (the number of O-type stars is particularly poorly known for many of these regions). Despite this some comparisons may be made. Cyg~OB2 is certainly a very high-mass cluster, comparable to the Arches and Quintuplet clusters near the Galactic center or the giant H~{\sc ii} regions Westerlund~2 and NGC~3603 in the Carina spiral arm. Only regions such as Westerlund~1 cluster, W49A, and R136 in the Large Magellanic Cloud appear to be more massive.

\begin{table*}
\begin{center}
\caption{Comparison of the properties of Cyg~OB2 with other regions. The number of O-type stars for Cyg~OB2 are only those spectroscopically confirmed, while for other regions they are generally extrapolated from photometry.}
\label{msfrs}
\begin{tabular}{@{}l ccccc}
\hline
Star forming		& Age		& $M_{total}$		& Core density		& Core radius	& O stars	\\
region			& (Myrs)		& ($M_{\odot}$)	& ($M_{\odot}$ pc$^{-3}$) & (pc) & \\
\hline
ONC $^a$		& 0.3--1.0 & $1.9 \times 10^3$	& $2 \times 10^4$ 		& 0.16-0.21	& 1 \\
Westerlund 2 $^b$	& $2.0 \pm 0.3$& $7.0 \times 10^3$ 	& -				& 0.002-0.2	& $> 24$\\
Quintuplet	 Cluster$^c$& $4.0 \pm 2.0$& 6--10~$\times 10^3$ 	&  $2.5 \times 10^2$	& 1.0		& -\\	
NGC 3603 $^d$	& $1.0 \pm 0.35$& $1.0 \times 10^4$& $1 \times 10^5$	& 0.23		&  $\sim$40\\
Arches Cluster $^e$	& $2.5 \pm 0.5$& $> 1 \times 10^4$	& $3 \times 10^5$ 	& 0.034-0.30	& 120-160\\
Cygnus OB2		& $\sim$1--7 & $3 \times 10^4$ 	& $1 \times 10^2$	& 1.3			& 65 \\
Westerlund 1 $^f$	& $3.6 \pm 0.7$& $5 \times 10^4$ 	& $>3.6 \times 10^3$& 0.4		& $> 50$\\
W49A $^g$		& $1.0 \pm 1.0$ & $6 \times 10^4$	& - 				& -			& $> 100$\\
R136 $^g$		& $3.0 \pm 2.0$& $>1 \times 10^5$ 	& -				& -			& $> 150$\\
\hline
\end{tabular}
\\
~\\
References:
$^a$ \citet{hill97} and \citet{hill98b} ;
$^b$ \citet{asce07} and \citet{naze08} ;
$^c$ \citet{fige99a} and \citet{fige99b} ;
$^d$ \citet{grab88}, \citet{dris95}, and \citet{stol06} ;
$^e$ \citet{moff94}, \citet{fige99b}, \citet{fige02}, \citet{stol02} ;
$^f$ \citet{bran08} ;
$^g$ \citet{cont02}, \citet{alve03}, and \citet{home05} ;
$^h$ \citet{mass98} and \citet{bran07}.
\end{center}
\end{table*}

Despite their similar masses, these SFRs show significant differences in their stellar densities, ranging from $\sim10^2$~$M_{\odot}$ pc$^{-3}$ for Cyg~OB2 and the Quintuplet cluster to $\sim10^5$~$M_{\odot}$ pc$^{-3}$ for NGC~3603 and the Arches cluster. This may represent an inherent difference between two varieties of massive star forming region: that of the compact, massive cluster with a very high density core on one hand and the lower density and spatially more extended region on the other. This could be an evolutionary effect since Cyg~OB2 and the Quintuplet cluster are both older than NGC~3603 and the Arches cluster, although the age differences are much shorter than the time periods necessary for such a change \citep{giel08}. It is more likely that these differences are real and are related to how the molecular cloud collapsed: low density cores suggest formation by triggering or the interaction of gas motions, while gravitational collapse typically results in higher densities \citep{smit08a}.

\citet{home05} note that W49A, which also has a low core density, has a very subclustered structure, a picture that could apply to the wider Cygnus~X giant molecular cloud (GMC) with subclusters such as Cyg~OB2 \citep{bica03} and Cyg~OB9 \citep[e.g.][]{uyan01}. This structure supports the hierarchical picture of star formation where young clusters and associations represent the dense inner regions of a hierarchy emerging from the GMC in which they formed \citep[e.g.][]{elme06}.

\section{Conclusions}

We have analysed the integrated stellar properties of young stars in the massive SFR Cygnus~OB2 using the recently published catalogue of {\it Chandra} X-ray point sources detected in the region. We find that our sample is complete down to $\sim$1~M$_{\odot}$, excluding A and B-type stars that are not expected to emit X-rays, making this one of the deepest surveys of the region. 
Ages for sources in the two fields studied were estimated from pre-MS isochrone fits to the near-IR CMD and found to be $3.5^{+0.75}_{-1.0}$ and $5.25^{+1.5}_{-1.0}$~Myrs, both with significant spreads around the isochrones that cannot solely be explained by photometric errors. 
Given that the most massive stars in the region have an age of $2 \pm 1$~Myrs \citep{mass91,hans03}, we suggest that this supports recent evidence for an older generation of stars in the region \citep[e.g.][]{drew08,come08} that either represent a significant age spread or multiple epochs of star formation. 
Additionally, the lack of sizable population of highly embedded sources suggests that star formation in the region studied has declined significantly. 

Assuming that a significant fraction of sources in Cyg~OB2 are older than the previously accepted age of 2~Myrs, many of the unique properties of the region may be explained. As noted by \citet{drew08}, the distinct lack of a bright H{\sc ii} region surrounding Cyg~OB2 is perfectly reasonable if the clearing timescale has been greater than 5~Myrs. Furthermore, we propose that the low fraction of sources with $K$-band excesses found by \citet[][and confirmed from our larger sample]{alba07} is less abnormal if a significant fraction of sources are older than previously believed.

We measure the stellar mass functions for stars within our completeness limits and find a slope of $\Gamma = -1.09 \pm 0.13$, in good agreement with the potentially universal value estimated by \citet{krou02}. This is the deepest and most accurate measurement of the mass function in Cyg~OB2, building on previously measurements that have found either steeper \citep{knod00} or shallower \citep{mass91} slopes. A steepening of the mass function is observed at high masses, as found by \citet{kimi07}, which we attribute to the demise of the highest mass members of the older generations in the region.

An estimation of the total size of the region suggests that the stellar density drops off faster than estimated by \citet{knod00}. This, combined with our mass function estimate, reveals a total mass of $\sim 3 \times 10^4$~M$_{\odot}$. This makes Cyg~OB2 similar in mass to many of the recently discovered massive clusters in our Galaxy such as the Arches and Quintuplet clusters. However, we note that comparing other properties of these massive star forming regions, such as their central densities, reveals differences that may hint at different formation mechanisms. 

Past studies of Cyg~OB2 have clearly been complicated by an uncertain star formation history and a large spatial extent in a crowded region of the Galactic Plane. These are likely to be inherent properties of massive star forming regions in galactic spiral arms where star formation continues for longer periods and across larger areas, propagated through entire GMCs by continued rejuvenation and feedback triggering. These findings support the need to fully understand the full star formation history of a region before if its observed properties are to be correctly interpreted. Cygnus~OB2 is clearly a very important region to study, not only because of its proximity and size, but also because of the insights it will give us into galactic-scale star formation and the evolution of GMCs.

\acknowledgments

We thank Phil Lucas and Yvonne Unruh for careful reading of this paper and helpful discussions. We also thank the anonymous referee for helpful suggestions that have improved this work. 
This research has made use of data from the {\it Chandra X-ray Observatory} (operated by the Smithsonian Astrophysical Observatory on behalf of NASA) obtained from the {\it Chandra} Data Archive. 
This publication makes use of data products from IPHAS (carried out at the Isaac Newton Telescope, INT, operated on the island of La Palma by the Isaac Newton Group), 2MASS (a joint project of the University of Massachusetts and the Infrared Processing and Analysis Center / California Institute of Technology, funded by NASA and the NSF), and UKIDSS (supported by the UKATC and CASU). 
JJD was funded by NASA contract NAS8-39073 to the {\it Chandra X-ray Center} (CXC) during the course of this research and thanks the CXC director, Harvey Tananbaum, and the science team for advice and support. 
NJW acknowledges an SAO Pre-doctoral Fellowship.

\bibliography{/Users/nick/Documents/Work/tex_papers/bibliography.bib}

\begin{thebibliography}{87}
\expandafter\ifx\csname natexlab\endcsname\relax\def\natexlab#1{#1}\fi

\bibitem[{{Abdo} {et~al.}(2009){Abdo}, et al.}]{abdo09}
{Abdo}, A.~A., et al., 2009, Science, 325, 840

\bibitem[{{Albacete Colombo} {et~al.}(2007){Albacete Colombo}, {Flaccomio},
  {Micela}, {Sciortino}, \& {Damiani}}]{alba07}
{Albacete Colombo}, J.~F., {Flaccomio}, E., {Micela}, G., {Sciortino}, S., \&
  {Damiani}, F. 2007, \aap, 464, 211

\bibitem[{{Alves} \& {Homeier}(2003)}]{alve03}
{Alves}, J., \& {Homeier}, N. 2003, \apjl, 589, L45

\bibitem[{{Ascenso} {et~al.}(2007){Ascenso}, {Alves}, {Beletsky}, \&
  {Lago}}]{asce07}
{Ascenso}, J., {Alves}, J., {Beletsky}, Y., \& {Lago}, M.~T.~V.~T. 2007, \aap,
  466, 137

\bibitem[{{Balog} {et~al.}(2007){Balog}, {Muzerolle}, {Rieke}, {Su}, {Young},
  \& {Megeath}}]{balo07}
{Balog}, Z., {Muzerolle}, J., {Rieke}, G.~H., {Su}, K.~Y.~L., {Young}, E.~T.,
  \& {Megeath}, S.~T. 2007, \apj, 660, 1532

\bibitem[{{Baraffe} {et~al.}(2009){Baraffe}, {Chabrier}, \&  {Gallardo}}]{bara09}
{Baraffe}, I., {Chabrier}, G., \& {Gallardo}, J. 2009, \apjl, 702, L27

\bibitem[{{Bednarek}(2003)}]{bedn03}
{Bednarek}, W. 2003, \mnras, 345, 847

\bibitem[{{Berghoefer} {et~al.}(1997){Berghoefer}, {Schmitt}, {Danner}, \&
  {Cassinelli}}]{berg97}
{Berghoefer}, T.~W., {Schmitt}, J.~H.~M.~M., {Danner}, R., \& {Cassinelli},
  J.~P. 1997, \aap, 322, 167

\bibitem[{{Bessell}(2005)}]{bess05}
{Bessell}, M.~S. 2005, \araa, 43, 293

\bibitem[{{Bica} {et~al.}(2003){bica}, {Bonatto}, \& {Dutra}}]{bica03}
{Bica}, E., {Bonatto}, C., \& {Dutra}, C.~M. 2003, \aap, 405, 991

\bibitem[{{Brandl} {et~al.}(2007){Brandl}, {Portegies Zwart}, {Moffat}, \&
  {Chernoff}}]{bran07}
{Brandl}, B.~R., {Portegies Zwart}, S.~F., {Moffat}, A.~F.~J., \& {Chernoff},
  D.~F. 2007, in Astronomical Society of the Pacific Conference Series, Vol.
  367, Massive Stars in Interactive Binaries, ed. N.~{St.-Louis} \& A.~F.~J.
  {Moffat}, 629

\bibitem[{{Brandner} {et~al.}(2008){Brandner}, {Clark}, {Stolte}, {Waters},
  {Negueruela}, \& {Goodwin}}]{bran08}
{Brandner}, W., {Clark}, J.~S., {Stolte}, A., {Waters}, R., {Negueruela}, I.,
  \& {Goodwin}, S.~P. 2008, \aap, 478, 137

\bibitem[{{Broos} {et~al.}(2007){Broos}, {Feigelson}, {Townsley}, {Getman},
  {Wang}, {Garmire}, {Jiang}, \& {Tsuboi}}]{broo07}
{Broos}, P.~S., {Feigelson}, E.~D., {Townsley}, L.~K., {Getman}, K.~V., {Wang},
  J., {Garmire}, G.~P., {Jiang}, Z., \& {Tsuboi}, Y. 2007, \apjs, 169, 353

\bibitem[{{Camilo} {et~al.}(2009){Camilo}, et al.}]{cami09}
{Camilo}, F., et al.,  2009, \apj, 705, 1

\bibitem[{{Comer{\'o}n} \& {Pasquali}(2007)}]{come07}
{Comer{\'o}n}, F., \& {Pasquali}, A. 2007, \aap, 467, L23

\bibitem[{{Comer{\'o}n} {et~al.}(2008){Comer{\'o}n}, {Pasquali}, {Figueras}, \&
  {Torra}}]{come08}
{Comer{\'o}n}, F., {Pasquali}, A., {Figueras}, F., \& {Torra}, J. 2008, \aap,
  486, 453

\bibitem[{{Comer{\'o}n} {et~al.}(2002){Comer{\'o}n}, et al.}]{come02}
{Comer{\'o}n}, F., et al., 2002, \aap, 389, 874

\bibitem[{{Conti} \& {Blum}(2002)}]{cont02}
{Conti}, P.~S., \& {Blum}, R.~D. 2002, \apj, 564, 827

\bibitem[{{Cramer}(1997)}]{cram97}
{Cramer}, N. 1997, in ESA Special Publication, Vol. 402, Hipparcos - Venice
  '97, 311--314

\bibitem[{{Drew} {et~al.}(2005){Drew}, et al.}]{drew05}
{Drew}, J.~E., et al., 2005,  \mnras, 362, 753

\bibitem[{{Drew} {et~al.}(2008){Drew}, {Greimel}, {Irwin}, \& {Sale}}]{drew08}
{Drew}, J.~E., {Greimel}, R., {Irwin}, M.~J., \& {Sale}, S.~E. 2008, \mnras,
  386, 1761

\bibitem[{{Drissen} {et~al.}(1995){Drissen}, {Moffat}, {Walborn}, \& {Shara}}]{dris95}
{Drissen}, L., {Moffat}, A.~F.~J., {Walborn}, N.~R., \& {Shara}, M.~M. 1995,  \aj, 110, 2235

\bibitem[{{Elmegreen} {et~al.}(2006){Elmegreen}, {Elmegreen}, {Chandar},
  {Whitmore}, \& {Regan}}]{elme06}
{Elmegreen}, B.~G., {Elmegreen}, D.~M., {Chandar}, R., {Whitmore}, B., \&
  {Regan}, M. 2006, \apj, 644, 879

\bibitem[{{Feigelson} {et~al.}(2005){Feigelson}, et al.}]{feig05}
{Feigelson}, E.~D., et al., 2005,  \apjs, 160, 379

\bibitem[{{Figer} {et~al.}(1999{\natexlab{a}}){Figer}, {Kim}, {Morris},
  {Serabyn}, {Rich}, \& {McLean}}]{fige99a}
{Figer}, D.~F., {Kim}, S.~S., {Morris}, M., {Serabyn}, E., {Rich}, R.~M., \&
  {McLean}, I.~S. 1999{\natexlab{a}}, \apj, 525, 750

\bibitem[{{Figer} {et~al.}(1999{\natexlab{b}}){Figer}, {McLean}, \&  {Morris}}]{fige99b}
{Figer}, D.~F., {McLean}, I.~S., \& {Morris}, M. 1999{\natexlab{b}}, \apj, 514,  202

\bibitem[{{Figer} {et~al.}(2002){Figer}, et al.}]{fige02}
{Figer}, D.~F., et al. 2002, \apj, 581, 258

\bibitem[{{Getman} {et~al.}(2005{\natexlab{a}}){Getman}, {Feigelson}, {Grosso},
  {McCaughrean}, {Micela}, {Broos}, {Garmire}, \& {Townsley}}]{getm05}
{Getman}, K.~V., {Feigelson}, E.~D., {Grosso}, N., {McCaughrean}, M.~J.,
  {Micela}, G., {Broos}, P., {Garmire}, G., \& {Townsley}, L.
  2005{\natexlab{a}}, \apjs, 160, 353

\bibitem[{{Getman} {et~al.}(2006){Getman}, {Feigelson}, {Townsley}, {Broos},
  {Garmire}, \& {Tsujimoto}}]{getm06}
{Getman}, K.~V., {Feigelson}, E.~D., {Townsley}, L., {Broos}, P., {Garmire},
  G., \& {Tsujimoto}, M. 2006, \apjs, 163, 306

\bibitem[{{Getman} {et~al.}(2005{\natexlab{b}}){Getman}, et al.}]{getm05a}
{Getman}, K.~V., et al., 2005{\natexlab{b}}, \apjs, 160, 319

\bibitem[{{Gieles} \& {Bastian}(2008)}]{giel08}
{Gieles}, M., \& {Bastian}, N. 2008, \aap, 482, 165

\bibitem[{{Gorenstein}(1975)}]{gore75}
{Gorenstein}, P. 1975, \apj, 198, 95

\bibitem[{{Grabelsky} {et~al.}(1988){Grabelsky}, {Cohen}, {Bronfman}, \& {Thaddeus}}]{grab88}
{Grabelsky}, D.~A., {Cohen}, R.~S., {Bronfman}, L., \& {Thaddeus}, P. 1988,  \apj, 331, 181

\bibitem[{{Guarcello} {et~al.}(2007){Guarcello}, {Prisinzano}, {Micela},
  {Damiani}, {Peres}, \& {Sciortino}}]{guar07}
{Guarcello}, M.~G., {Prisinzano}, L., {Micela}, G., {Damiani}, F., {Peres}, G.,
  \& {Sciortino}, S. 2007, \aap, 462, 245

\bibitem[{{Haisch} {et~al.}(2001{\natexlab{a}}){Haisch}, {Lada}, \& {Lada}}]{hais01a}
{Haisch}, Jr., K.~E., {Lada}, E.~A., \& {Lada}, C.~J. 2001{\natexlab{a}}, \aj,  121, 2065

\bibitem[{{Hanson}(2003)}]{hans03}
{Hanson}, M.~M. 2003, \apj, 597, 957

\bibitem[{{Herbst} {et~al.}(1994){Herbst}, {Herbst}, {Grossman}, \& {Weinstein}}]{herb94}
{Herbst}, W., {Herbst}, D.~K., {Grossman}, E.~J., \& {Weinstein}, D. 1994, \aj, 108, 1906

\bibitem[{{Hewett} {et~al.}(2006){Hewett}, {Warren}, {Leggett}, \&  {Hodgkin}}]{hewe06}
{Hewett}, P.~C., {Warren}, S.~J., {Leggett}, S.~K., \& {Hodgkin}, S.~T. 2006,  \mnras, 367, 454

\bibitem[{{Hillenbrand}(1997)}]{hill97}
{Hillenbrand}, L.~A. 1997, \aj, 113, 1733

\bibitem[{{Hillenbrand}(2005)}]{hill05}
{Hillenbrand}, L.~A. 2005, in {STScI Symposium Series}, Vol.~19, {A Decade of
  Discovery: Planets Around Other Stars}, ed. M.~{Livio}

\bibitem[{{Hillenbrand} \& {Hartmann}(1998)}]{hill98b}
{Hillenbrand}, L.~A., \& {Hartmann}, L.~W. 1998, \apj, 492, 540

\bibitem[{{Hillenbrand} {et~al.}(1998){Hillenbrand}, {Strom}, {Calvet},
  {Merrill}, {Gatley}, {Makidon}, {Meyer}, \& {Skrutskie}}]{hill98}
{Hillenbrand}, L.~A., {Strom}, S.~E., {Calvet}, N., {Merrill}, K.~M., {Gatley},
  I., {Makidon}, R.~B., {Meyer}, M.~R., \& {Skrutskie}, M.~F. 1998, \aj, 116,  1816

\bibitem[{{Hillenbrand} \& {White}(2004)}]{hill04}
{Hillenbrand}, L.~A., \& {White}, R.~J. 2004, \apj, 604, 741

\bibitem[{{Hoffmeister} {et~al.}(2008){Hoffmeister}, {Chini}, {Scheyda},
  {Schulze}, {Watermann}, {N{\"u}rnberger}, \& {Vogt}}]{hoff08}
{Hoffmeister}, V.~H., {Chini}, R., {Scheyda}, C.~M., {Schulze}, D.,
  {Watermann}, R., {N{\"u}rnberger}, D., \& {Vogt}, N. 2008, \apj, 686, 310

\bibitem[{{Homeier} \& {Alves}(2005)}]{home05}
{Homeier}, N.~L., \& {Alves}, J. 2005, \aap, 430, 481

\bibitem[{{Howarth}(1983)}]{howa83}
{Howarth}, I.~D. 1983, \mnras, 203, 301

\bibitem[{{Johnstone} {et~al.}(1998){Johnstone}, {Hollenbach}, \& {Bally}}]{john98}
{Johnstone}, D., {Hollenbach}, D., \& {Bally}, J. 1998, \apj, 499, 758

\bibitem[{{Kenyon} \& {Hartmann}(1995)}]{keny95}
{Kenyon}, S.~J., \& {Hartmann}, L. 1995, \apjs, 101, 117

\bibitem[{{Kiminki} {et~al.}(2007){Kiminki}, et al.}]{kimi07}
{Kiminki}, D.~C., et al., 2007, \apj, 664, 1102

\bibitem[{{King}(1962)}]{king62}
{King}, I. 1962, \aj, 67, 471

\bibitem[{{Kn{\"o}dlseder}(2000)}]{knod00}
{Kn{\"o}dlseder}, J. 2000, \aap, 360, 539

\bibitem[{{Koornneef}(1983)}]{koor83}
{Koornneef}, J. 1983, \aap, 128, 84

\bibitem[{{Kroupa}(2001)}]{krou01a}
{Kroupa}, P. 2001, \mnras, 322, 231

\bibitem[{{Kroupa}(2002)}]{krou02}
---. 2002, Science, 295, 82

\bibitem[{{Lada}(1987)}]{lada87}
{Lada}, C.~J. 1987, in IAU Symposium, Vol. 115, Star Forming Regions, ed. 
M.~{Peimbert} \& J.~{Jugaku}, 1--17

\bibitem[{{Lada} \& {Lada}(2003)}]{lada03}
{Lada}, C.~J., \& {Lada}, E.~A. 2003, \araa, 41, 57

\bibitem[{{Lucas} {et~al.}(2008){Lucas}, et al.}]{luca08}
{Lucas}, P.~W., et al., 2008, \mnras, 391, 136

\bibitem[{{Luhman} {et~al.}(2000){Luhman}, {Rieke}, {Young}, {Cotera}, {Chen},
  {Rieke}, {Schneider}, \& {Thompson}}]{luhm00}
{Luhman}, K.~L., {Rieke}, G.~H., {Young}, E.~T., {Cotera}, A.~S., {Chen}, H.,
  {Rieke}, M.~J., {Schneider}, G., \& {Thompson}, R.~I. 2000, \apj, 540, 1016

\bibitem[{{Martins} {et~al.}(2005){Martins}, {Schaerer}, \& {Hillier}}]{mart05}
{Martins}, F., {Schaerer}, D., \& {Hillier}, D.~J. 2005, \aap, 436, 1049

\bibitem[{{Massey}(1998)}]{mass98}
{Massey}, P. 1998, in Astronomical Society of the Pacific Conference Series,
  Vol. 142, The Stellar Initial Mass Function (38th Herstmonceux Conference),
  ed. G.~{Gilmore} \& D.~{Howell}, 17

\bibitem[{{Massey} \& {Thompson}(1991)}]{mass91}
{Massey}, P., \& {Thompson}, A.~B. 1991, \aj, 101, 1408

\bibitem[{{McCuskey} \& {Houk}(1971)}]{mccu71}
{McCuskey}, W.~W., \& {Houk}, N. 1971, \aj, 76, 1117

\bibitem[{{Meyer} {et~al.}(1997){Meyer}, {Calvet}, \& {Hillenbrand}}]{meye97}
{Meyer}, M.~R., {Calvet}, N., \& {Hillenbrand}, L.~A. 1997, \aj, 114, 288

\bibitem[{{Moffat} {et~al.}(1994){Moffat}, {Drissen}, \& {Shara}}]{moff94}
{Moffat}, A.~F.~J., {Drissen}, L., \& {Shara}, M.~M. 1994, \apj, 436, 183

\bibitem[{{Muench} {et~al.}(2008){Muench}, {Getman}, {Hillenbrand}, \&
  {Preibisch}}]{muen08}
{Muench}, A., {Getman}, K., {Hillenbrand}, L., \& {Preibisch}, T. 2008, {Star
  Formation in the Orion Nebula I: Stellar Content}, ed. {Reipurth, B.}, 483--+

\bibitem[{{Naz{\'e}} {et~al.}(2008){Naz{\'e}}, {Rauw}, \& {Manfroid}}]{naze08}
{Naz{\'e}}, Y., {Rauw}, G., \& {Manfroid}, J. 2008, \aap, 483, 171

\bibitem[{{Negueruela} {et~al.}(2008){Negueruela}, {Marco}, {Herrero}, \&  {Clark}}]{negu08}
{Negueruela}, I., {Marco}, A., {Herrero}, A., \& {Clark}, J.~S. 2008, \aap,  487, 575

\bibitem[{{Palla} \& {Stahler}(2000)}]{pall00}
{Palla}, F., \& {Stahler}, S.~W. 2000, \apj, 540, 255

\bibitem[{{Pallavicini} {et~al.}(1981){Pallavicini}, {Golub}, {Rosner},
  {Vaiana}, {Ayres}, \& {Linsky}}]{pall81}
{Pallavicini}, R., {Golub}, L., {Rosner}, R., {Vaiana}, G.~S., {Ayres}, T., \&
  {Linsky}, J.~L. 1981, \apj, 248, 279

\bibitem[{{Preibisch} \& {Feigelson}(2005)}]{prei05}
{Preibisch}, T., \& {Feigelson}, E.~D. 2005, \apjs, 160, 390

\bibitem[{{Preibisch} {et~al.}(2005){Preibisch}, et al.}]{prei05a}
{Preibisch}, T., et al., 2005, \apjs, 160, 401

\bibitem[{{Preibisch} \& {Zinnecker}(1999)}]{prei99}
{Preibisch}, T., \& {Zinnecker}, H. 1999, \aj, 117, 2381

\bibitem[{{Reddish} {et~al.}(1967){Reddish}, {Lawrence}, \& {Pratt}}]{redd67}
{Reddish}, V.~C., {Lawrence}, L.~C., \& {Pratt}, N.~M. 1967, {PROE}, 5

\bibitem[{{Ryter}(1996)}]{ryte96}
{Ryter}, C.~E. 1996, \apss, 236, 285

\bibitem[{{Sale} {et~al.}(2009){Sale}, et al.}]{sale09}
{Sale}, S.~E., et al., 2009, \mnras, 392, 497

\bibitem[{{Salpeter}(1955)}]{salp55}
{Salpeter}, E.~E. 1955, \apj, 121, 161

\bibitem[{{Scalo}(1986)}]{scal86}
{Scalo}, J.~M. 1986, Fundamentals of Cosmic Physics, 11, 1

\bibitem[{{Schlegel} {et~al.}(1998){Schlegel}, {Finkbeiner}, \&  {Davis}}]{schl98}
{Schlegel}, D.~J., {Finkbeiner}, D.~P., \& {Davis}, M. 1998, \apj, 500, 525

\bibitem[{{Siess} {et~al.}(2000){Siess}, {Dufour}, \& {Forestini}}]{sies00}
{Siess}, L., {Dufour}, E., \& {Forestini}, M. 2000, \aap, 358, 593

\bibitem[{{Skrutskie} {et~al.}(2006){Skrutskie}, {Cutri}, {Stiening},
  {Weinberg}, {Schneider}, {Carpenter}, \& {Beichman}}]{skru06}
{Skrutskie}, M.~F., {Cutri}, R.~M., {Stiening}, R., {Weinberg}, M.~D.,
  {Schneider}, S., {Carpenter}, J.~M., \& {Beichman}, C. 2006, \aj, 131, 1163

\bibitem[{{Smith}(2008)}]{smit08a}
{Smith}, L.~J. 2008, in Astronomical Society of the Pacific Conference Series,  Vol. 390, 
  Pathways Through an Eclectic Universe, ed. {J.~H.~Knapen, T.~J.~Mahoney, \& A.~Vazdekis}, 39

\bibitem[{{Smith} \& {Gallagher}(2001)}]{smit01a}
{Smith}, L.~J., \& {Gallagher}, J.~S. 2001, \mnras, 326, 1027

\bibitem[{{Stelzer} {et~al.}(2006){Stelzer}, {Hu{\'e}lamo}, {Micela}, \& {Hubrig}}]{stel06}
{Stelzer}, B., {Hu{\'e}lamo}, N., {Micela}, G., \& {Hubrig}, S. 2006, \aap, 452, 1001

\bibitem[{{Stolte} {et~al.}(2006){Stolte}, {Brandner}, {Brandl}, \&  {Zinnecker}}]{stol06}
{Stolte}, A., {Brandner}, W., {Brandl}, B., \& {Zinnecker}, H. 2006, \aj, 132, 253

\bibitem[{{Stolte} {et~al.}(2002){Stolte}, {Grebel}, {Brandner}, \& {Figer}}]{stol02}
{Stolte}, A., {Grebel}, E.~K., {Brandner}, W., \& {Figer}, D.~F. 2002, \aap, 394, 459

\bibitem[{{Uyan{\i}ker} {et~al.}(2001){Uyan{\i}ker}, {F{\"u}rst}, {Reich},
 {Aschenbach}, \& {Wielebinski}}]{uyan01}
{Uyan{\i}ker}, B., {F{\"u}rst}, E., {Reich}, W., {Aschenbach}, B., \&
  {Wielebinski}, R. 2001, \aap, 371, 675

\bibitem[{{Vaiana} {et~al.}(1981){Vaiana}, et al.}]{vaia81}
{Vaiana}, G.~S., et al., 1981, \apj, 245, 163

\bibitem[{{Vink} {et~al.}(2008){Vink}, {Drew}, {Steeghs}, {Wright}, {Martin},
  {G{\"a}nsicke}, {Greimel}, \& {Drake}}]{vink08}
{Vink}, J.~S., {Drew}, J.~E., {Steeghs}, D., {Wright}, N.~J., {Martin}, E.~L.,
  {G{\"a}nsicke}, B.~T., {Greimel}, R., \& {Drake}, J. 2008, \mnras, 387, 308

\bibitem[{{Wang} {et~al.}(2007){Wang}, {Townsley}, {Feigelson}, {Getman},
  {Broos}, {Garmire}, \& {Tsujimoto}}]{wang07}
{Wang}, J., {Townsley}, L.~K., {Feigelson}, E.~D., {Getman}, K.~V., {Broos},
  P.~S., {Garmire}, G.~P., \& {Tsujimoto}, M. 2007, \apjs, 168, 100

\bibitem[{{Wright} \& {Drake}(2009)}]{wrig09a}
{Wright}, N.~J., \& {Drake}, J.~J. 2009, \apjs, 184, 84

\end{thebibliography}

\end{document}